Title: Carbon diffusion in α-iron: Evidence for quantum mechanical tunneling
Authors: Ludwik Dabrowski (1), Alexander Andreev (2), Mladen Georgiev (2) ((1) Institute Atomic Energy, Otwock-Swierk, Poland, (2) Institute of Solid State Physics, Bulgarian Academy of Sciences, Sofia, Bulgaria)
Comments: 12 pages, 3 figures, 1 table; all pdf format
Journal-ref: Under review in Metallurgical and Materials Transactions

Recent experimental data on the diffusion coefficient of carbon in α-iron below liquid nitrogen temperature (LNT) question the classical approach to the observed temperature dependence. As the temperature is lowered below LNT, the diffusion constant tends to a nearly temperature-independent value rather than continuing its activated trend. The low temperature branch is apparently characteristic of a quantum mechanical process dominated by tunneling in ground state. Concomitantly we apply an occurrence-probability approach to describing the overall temperature dependence as a single continuous rate. Within the adiabatic approximation the electronic eigenvalue depending parametrically on the nuclear coordinates is taken to be the potential energy to control the motion of the nuclei. The resulting rate involves all horizontal-tunneling energy-conserving elastic transitions at the quantized energy levels of the migrating atom. A small though not negligible slope in the temperature dependence as the temperature is raised below 100 K is dealt with by complementing for the rate of a parallel one-phonon inelastic-tunneling process in excess to the basic elastic-tunneling rate. Our combined approach agrees well with the experimental data. In particular, the frequency of the coupled vibration is obtained virtually identical to the carbon vibrational frequency from inelastic neutron scattering data. The migrational barrier is also found to be within the limits expected for α-iron.

1. Introduction

Carbon diffusion in iron is among the best known cases of impurity migration in solids. For great many years it has provided what is believed to be a textbook example for the Arrhenius temperature dependence of a diffusion coefficient

$$D(T) = D_0 \exp(-E_m / k_B T) \qquad (1)$$

extending over 7 decades of magnitude [1÷4]. Here $D_0$ is the frequency factor and $E_m$ is the migration barrier. For a reference, see Fig. 4.7 in Wert & Thomson's monograph where $D_0 = 2 \times 10^{-5}$ m$^2$/s and $E_m = 0.9$ eV. It has also been taken for granted that the diffusion mechanism involves classic jumps across well-defined interstitial sites in the bcc iron lattice. If so, the whole thermal activation would go to overcoming the migration barrier rather than splitting it nearly equally between migration and vacancy formation.

The classic theory leads to

$$D_0 = f(\zeta/6) d^2 \nu \qquad (2)$$

where $\zeta$ is the coordination number at the jump site (e.g. $\zeta = 4$ for a vacancy jump in the bcc lattice), d is the jump distance and $\nu$ is the attempt frequency of the vibrating atom. The quantity $k_m(T) = \nu \exp(-E_m/k_BT)$ is the classic transition rate, i.e. the number of overbarrier jumps per unit time. The factor f accounts for the probability that the migrating atom does not jump back to the original site after making a jump forward.

We note in passing that a classical frequency factor $D_0 = 2\times10^{-5}$ m$^2$/s implies that the vibrational frequency coupled to the diffusing carbon in iron is in excess of most vibrational frequencies observed so far in poorly conducting solids. Indeed with d = 2.48 Å (the nearest-neighbor separation in $\alpha$-iron) and $\zeta = 4$ we get $f\nu = 4.88\times10^{14}$ s$^{-1}$. Even with $f \sim 1$ we arrive at $\hbar\omega \sim 2$ eV ! We conclude that the frequency factor $D_0$ may not be accounted for by the classic theory.

Nevertheless, as our own analysis below suggests carbon diffusion is likely to occur via the tetrahedral sites. This changes the effective squared jump distance $d^2$ by a factor of 12 which lowers the attempt frequency to 1360 cm$^{-1}$, close to the Raman frequency of carbon in diamond. We thank the referee for his essential comment at this point.

Yet, the agreement between classic theory and experiment has recently been found essentially questionable [5]. As the measurement range has been extended down to lower temperatures beyond the classic range, it has become clear that the carbon difussion coefficient at 4% C in $\alpha$-iron actually bends into another branch in which it is very weakly dependent on the temperature, if at all. The apparent activation energy of the lower temperature branch below 78 K being so low (< 2 meV), it is hardly attributable to any reasonable migration hindering energy barrier so that the branch may eventually be assumed temperature independent within the error bars. Alternatively then, the overall temperature dependence of the carbon diffusion coefficient will look like this: As the temperature is increased from zero-point, an almost constant lower temperature branch will be observed followed by a more or less gradual transition to a higher temperature Arrhenius portion. It will be characteristic of a migration through barrier tunneling at the lower temperatures followed by migration through overbarrier jumps at the higher temperatures.

In what comes next we will give arguments for a quantum mechanical extension of the theory and compare it with available experimental diffusion data. An option is provided by the small polaron theories based on the Born-Oppenheimer (B-O) approximation which assumes that locally the electrons follow adiabatically the nuclear motion. However, care will have to be taken to present the basic assumptions concisely though cautiously, since so far metals have hardly been the traditional testing ground for the quantal small polaron theories developed originally for poorly conducting solids. In this respect, we point to an earlier study applying a small polaron theory to the migration of positrons in metallic solids [6]. Foundations of the general theory of quantum diffusion in solids have apparently been laid down during the late seventies [7], albeit from a different angle, as will be seen shortly.

Other examples are provided by theoretical approaches to the diffusion of light interstitials in metals coupled to lattice vibrations [8]. The dynamics of a system (metal & diffusing particle) is decoupled into a fast dynamics of the light interstitial, which determines the "*bare*" tunneling rate, and the slower dynamics of the host atoms, the phonon *heat bath*, which couples to "*dress*" the tunneling rate [8]. There are several basic assumptions: First, the B-O approximation which separates electronic from nuclear motions. Second, the adiabatic approximation whereby a light interstitial follows the heat bath atoms adiabatically. Third, *Condon*'s approximation which states that the bare tunneling probability is independent of the phonon state, i.e. the temperature T. Fourth, the diffusing particle in a metal couples to the conduction electrons nonadiabatically resulting in a declining tunneling rate as the temperature is raised at low T.

## 2. Quantum mechanical migration

In quantum mechanical extensions the *bare rate* is defined one way or the other so as to incorporate tunneling transitions across the barrier. Examples can be found based on the B-O approximation by Fermi's Golden Rule to the multiphonon transitions, elastic (else phonon diagonal) or inelastic (else phonon off-diagonal), and by the occurrence probability approaches to the horizontal-tunneling elastic transitions [9,10]. We will presently follow Christov's two-site definition of the elastic-tunneling rate, as discussed at length elsewhere [11]. While the multiphonon theory has widely been applied to developing the premises of quantum diffusion [7], the elastic-tunneling theory appears to have enjoyed less popularity [11]. Nevertheless, the occurrence-probability theory is simpler and leads to physically transparent results. One way or the other, an extensive multiphonon excursion will be left for a subsequent publication.

### 2.1. Two-site Hamiltonian

To B-O approximation which separates electronic and nuclear variables the electrostatic potential at the migrating atom's site is given by the adiabatic energy of an electronic state centered at that site. In the two-site problem we correspondingly select a basis set of two static electron states $|1>$ & $|2>$. These are physically equivalent electronic states centered at two neighboring migrational sites (two-level problem). We then define the following two-site Hamiltonian:

$$H = E \, ( \, | \, 1 > < 1 \, | + | \, 2 > < 2 \, | \, ) + V_{12} \, ( \, | \, 1 > < 2 \, | + | \, 2 > < 1 \, | \, ) +$$

$$GQ \, ( \, | \, 1 > < 1 \, | - | \, 2 > < 2 \, | \, ) + \tfrac{1}{2} \, M\omega^2 Q^2 \qquad (3)$$

where *E* stands for the electron energy, $V_{12}$ is the electron energy splitting, *Q* is the mode coordinate, *G* is the electron-mode coupling constant. In (3) the first block of round brackets is the static electronic energy, the second one is the mixing energy of the electron states with each other and the third one is the electron-mode coupling energy. Mixing is essential for the migration transition from site 1 to site 2 to occur at all, for it splits the electronic terms thereby securing adiabaticity at the crossover. (This statement will become transparent shortly.) Ultimately the electron-mode coupling energy brings a temperature dependence to the diffusion coefficient. The out-of-phase coupling is typical for a displacement-promoting mode: one of the electronic states is squeezed, while the other one is extended. The last fourth term is the elastic energy of the migrating atom.

To work out an electronic potential for the migrating atom, we solve for Schrodinger's equation $H\psi = \varepsilon\psi$ by means of the linear combination $\psi = C_1 \, | \, 1 > + C_2 \, | \, 2 >$. Ultimately we derive the following roots of the secular equation:

$$\varepsilon_\pm(Q) = \tfrac{1}{2} \, \{(H_{11} + H_{22}) \pm \sqrt{[(H_{11} - H_{22})^2 + 4V_{12}^2]}\}$$

$$= \tfrac{1}{2} M\omega^2 Q^2 + E \pm \sqrt{[(GQ)^2 + V_{12}^2]} \qquad (4)$$

This is a double branch potential which controls the impurity atom migration along the active mode coordinate Q either in ground adiabatic state ($\varepsilon_-$) or in excited adiabatic state ($\varepsilon_+$). $\varepsilon_+(Q)$ and $\varepsilon_-(Q)$ avoid crossing at $Q = 0$ due to the off-diagonal splitting of $2V_{12}$. While the upper branch $\varepsilon_+(Q)$ is an anharmonic parabola bottomed at $Q = 0$, the lower branch $\varepsilon_-(Q)$ at $\eta < 1$ is composed of two isoenergy latteral wells, one on each side of the crossover coordinate, forming a barrier in-between (cf. Figure 3 below). Here and above $\eta = V_{12}/2\varepsilon_{CE}$ is the normalized interlevel energy gap. The lateral wells bottom at

$$Q_{0\pm} = \pm \sqrt{(G^4 - V_{12}^2 K^2)}/GK = \pm \sqrt{(2\varepsilon_{CE}/K)} \sqrt{(1-\eta^2)} \qquad (5)$$

where $K = M\omega^2$ is the stiffness. The interwell barrier at $Q = 0$ amounts to

$$\varepsilon_B = \varepsilon_-(0) - \varepsilon_-(Q_{0-}) = \varepsilon_{CE}(1-\eta)^2 \qquad (6)$$

while $\varepsilon_{CE} = G^2/2K$ is the electron-mode coupling energy.

## 2.2. Two-site rate

Christov's approach gives the two-site bare rate as a sum of weighted transition probabilities for horizontal (elastic) tunneling transitions at the quantized energy levels. Golden Rule alike, he assumes that the vibronic quantum states in the initial electronic state are in thermal equilibrium under Boltzmann statistics. He also applies Condon's approximation to factorize out the transition probabilities $W(E_n)$ into electronic $W_{el}(E_n)$ and nuclear $W_{nuke}(E_n)$ components. Under these conditions the quantal migration rate along the active mode coordinate of frequency $\nu$ reads:

$$\kappa_{mh}(T) = \nu (Z^\#/Z) \Sigma_{E(n)} W_{el}(E_n) W_{nuke}(E_n) \exp(-E_n/k_B T) \qquad (7)$$

where Z is the complete partition function and $Z^\#$ is the contribution to Z of all nonreactive modes. In all cases the electronic probabilities $W_{el}(E_n)$ are derived using an extension of Landau & Zener's formula for the (avoided) crossing of molecular terms [12], but accounting for the possiblity of multiple transitions forth and back [11]. The nuclear probabilities $W_{nuke}(E_n)$ are derived by a method due to Bardeen & Chrstov [10]:

$$W_{nuke}(E_n) = 4\pi^2 |U_{12}|^2 \sigma_1(E_n) \sigma_2(E_n)$$

$$U_{12} = (-\hbar^2/2M)[u_2^*(du_1/dq) - u_1(du_2^*/dq)]_{q=0} \qquad (8)$$

where $u_i$ is the wavefunction, $\sigma_i(E_n)$ is the DOS in either electronic state, $q$ is the scaled mode coordinate $q = (K/\hbar\omega)^{1/2} Q$, Q being the actual configurational coordinate.

Equation (8) gives the transition probabilities essentialy dependent on the form of the migration hindering barrier. We have considered two forms of a barrier: parabolic arising in an ensemble of linear harmonic oscillators and sinusoidal arising in an ensemble of nonlinear oscillators. Due to mathematical complexity, only the linear case will be considered in detail presently, while the nonlinear alternative will be left to a subsequent publication.

## 2.3. Parabolic barrier

Introducing the parabolic two-site potential (3) brings the nuclear problem to the familiar field of harmonic oscillators. Inserting $Z^{\#}/Z = 2\sinh(\hbar\omega/2k_BT)$ and $E_n = (n + \frac{1}{2})\hbar\omega$, we get for the two-site rate:

$$\kappa_{mh}(T) = 2\nu \sinh(\hbar\omega/2k_BT) \{\sum_{E(n)>>\epsilon(B)}\{2[1-\exp(-2\pi\gamma_n)]/[2-\exp(-2\pi\gamma_n)]\} \times$$

$$\{W_{nuke\ overbarrier}\} \exp(-E_n/k_BT) +$$

$$\sum_{E(n)<<\epsilon(B)}\{2\pi\gamma_n^{2\gamma(n)-1}\exp(-2\gamma_n) / [\Gamma(\gamma_n)]^2\}\{\pi[F_{nn}(q_0,q_C)/2^n n!]^2\exp(-\epsilon_R/\hbar\omega)\} \times$$

$$\exp(-E_n/k_BT)\} \qquad (9)$$

where the expressions put within the small curled brackets are the electron-transfer $W_{el}(E_n)$ and nuclear-tunneling $W_{nuke}(E_n)$ probabilities, overbarrier for $E_n > \epsilon_B$ and underbarrier for $E_n < \epsilon_B$, respectively,

$$\gamma_n = (\epsilon_{\alpha\beta}^2/8\hbar\omega)(\epsilon_R | E_n - \epsilon_C|)^{-1/2} \qquad (10)$$

is Landau-Zener's parameter,

$$F_{nn}(q_0,q_C) = 2q_0 H_n(q_C)H_n(q_C - 2q_0) - 2nH_{n-1}(q_C)H_{n-1}(q_C - 2q_0) +$$

$$2nH_n(q_C)H_{n-1}(q_C - 2q_0) \qquad (11)$$

is a quadratic form of Hermite polynomials, $\omega = 2\pi\nu$ is the angular vibrational frequency. We set $W_{nuke\ overbarrier} = 1$ for the overbarrier tunneling probability. $q_0$ and $q_C$ are the scaled well-bottom and crossover coordinates, respectively.

The remaining parameters are: $\epsilon_R$ - the lattice-reorganization energy, $\epsilon_C$ - the crossover energy, and $\omega_{bare}$ – the bare vibrational frequency. They relate to $\epsilon_B$, $\epsilon_{CE}$ and $\eta$ by way of

$$\epsilon_C = \epsilon_B(1+\eta^2)/(1-\eta)^2 = \epsilon_{CE}(1+\eta^2)$$
$$\epsilon_R = 4\epsilon_B(1+\eta)/(1-\eta) = 4\epsilon_{CE}(1-\eta^2) = 4\epsilon_C(1-\eta^2)/(1+\eta^2)$$
$$\omega_{bare} = \omega/\sqrt{(1-\eta^2)} \qquad (12)$$

The dynamic migration rate $\kappa_{mh}(T)$ depends on three fitting parameters; we presently take them to be $\eta$, $\epsilon_B$, and $\hbar\omega$. The isothermal symmetry of the two-site double well potential follows from the requirement that the migration steps should be reversible forth and back.

From the rate equation (9) it is easy to derive the clasical Arrhenius rate at temperatures sufficiently high to secure the complete predominance of the overbarrier transitions: As an illustration, from the first part of equation (9) using $W_{nuke} \sim 1$ and $\gamma_n \gg 1$ for overbarrier and adiabatic transitions, respectively, we get at $2k_BT \gg \hbar\omega$ (converting the sum into an integral by setting $dE_n \sim \hbar\omega$):

$$\kappa_{mh}(T) = \nu_{\varepsilon(B)} \int^{\infty} \exp(-E_n/k_BT) \, dE_n/k_BT = \nu \exp(-\varepsilon_B/k_BT)$$

### 3. Quantal diffusion coefficient

#### 3.1. Horizontal tunneling rate

Substituting the quantal migration rate $\kappa_{mh}(T)$ for the classic rate $k_m(T)$ we generalize equation (1) for the diffusion coefficient to get

$$D(T) = f(\zeta/6) \, d^2 \, \kappa_{mh}(T) \qquad (13)$$

It is essential to point out that the quantal rate $\kappa_{mh}(T)$ combines both the low temperature tunneling rate and the higher temperature classic rate as verified mathematically by means of equation (9). $\kappa_{mh}(T)$ summing up the partial rates at various vibronic energy levels which rates conserve the number of phonons, it gives the diffusion coefficient of a coherent migration process at any temperature. In particular, the total zero-point rate as obtained from equation (9) being

$$\kappa_{mh}(0) = (\varepsilon_R/\hbar) \exp(-\varepsilon_R/\hbar\omega), \qquad (14)$$

it results in a low temperature diffusion coefficient

$$D(0) = f(\zeta/6) \, d^2 (\varepsilon_R/\hbar) \exp(-\varepsilon_R/\hbar\omega) \qquad (15)$$

We see that displaying the quantal feature of carbon diffusion to a reasonable extent is largely due to the high coupled vibrational quantum $\hbar\omega$. From (14) and (2) we also get the relationship between the classical frequency factor and the quantal zero-point diffusion coefficient D(0):

$$D_0 = [D(0)/2\pi] (\hbar\omega/\varepsilon_R) \exp(\varepsilon_R/\hbar\omega) \qquad (16)$$

which possibly gives an alternative clue as to why the experimental frequency factor exceeds so largely the predicted classical $D_0$.

We shall further proceed by fitting equation (13) to the experimental data. The experimental diffusion coefficients are depicted by symbols in Figure 1. It can be seen that the low-temperature branch is succeeded from right to left by an almost flat plateau followed rather steeply by an Arrhenius branch at roughly $T_t \sim 250$ K. The abrupt transition to the thermally activated branch signifies the start of thermally populating the excited vibronic levels in an adiabatic process. From the bending point we estimate roughly the active mode frequency at $\hbar\omega \sim 4k_BT_t \sim 0.1$ eV. We also get $\varepsilon_B \sim 1$ eV from the slope of the thermally-activated branch. For the sake of simplicity, we assume that the electronic intersite transitions are all adiabatic, that is, we set $W_{el}(E_n) = 1$ at all $E_n$. Clearly, this is an oversimplification but it can help see where we are. We also find $\eta = 0.025$ to be the appropriate gap parameter. The dashed-line in Figure 1 depicts a good fit to the experimental points made by means of the elastic-tunneling isothermal theory of equation (13).

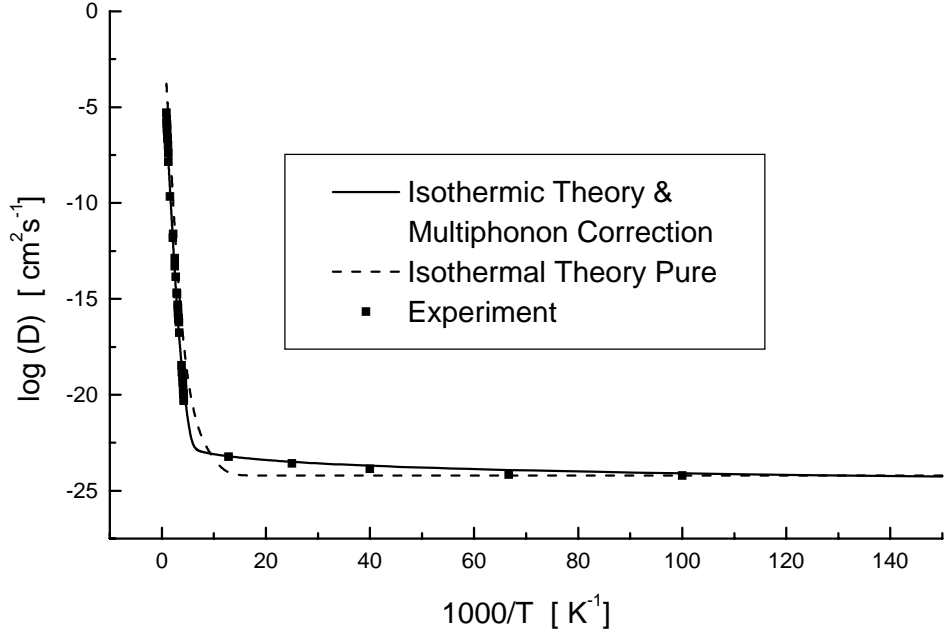

Figure 1. Temperature dependence of the diffusion coefficient of carbon in α-iron from Reference [2÷5] ($D$ in cm$^2$s$^{-1}$): experimental points by filled squares, fit of isothermal reaction-rate by a dashed line. A small difference between experiment and theory below LNT, due to one-phonon processes, is also accounted for by means of an additive correction of the form $AT$, where $A$ is constant and $T$ is temperature, leading to the solid line fit therein. $P = 6.5 \cdot 10^{-17}$, $A = 1.23 \cdot 10^{-9}$, $E_b = 1.1$ eV, $\hbar\omega = 0.077$ eV, $\eta = 0.025$.

### 3.2. Inelastic tunneling correction

We also see that the "flat plateau" in the temperature dependence between 10-100 K is in fact slowly and monotonically increasing as the temperature is raised. While the flat plateau is intrinsic of the horizontal (elastic) tunneling reaction rate, the monotonic rise is due to an inelastic-tunneling one-phonon absorption complementary to the basic elastic process. To account for the inelastic process, we add a rate term of a form derived for a phonon-coupled two-level system [13]:

$$\kappa_{mv} = C \coth (\Delta/k_B T) \qquad (17)$$

where $\Delta$ is the ground level vibronic tunneling splitting. Eq. (16) reduces to $\kappa_{mv} = C(k_B/\Delta)T \equiv AT$ at $T \gg \Delta/k_B$. For a typical value of $\Delta = 0.001$, this gives $T \gg 10$ K to cover most of the plateau range. We also estimate the relative weight of inelastic versus thermally-activated elastic transitions at the bending point $T_t$ to find that the former no longer make any sizeable share of the observed rate process. If eq.(17) is the only rate determining agent at low temperature, then

$$D(0) = f(\zeta/6) d^2 C \equiv f(\zeta/6) d^2 (\Delta/k_B) A \qquad (18)$$

In all other cases the zero-point diffusion coefficient will sum up of the right hand sides of eq.(15) and (17), that is,

$$D(0) = f(\zeta/6) d^2 [(\varepsilon_R/\hbar) \exp(-\varepsilon_R/\hbar\omega) + (\Delta/k_B) A] \quad (19)$$

In both equations (15) and (18) the scaling factor $P = f(\zeta/6) d^2$ is common, as it controls the conversion of rate to diffusion data.

A good fit to the experimental data as shown by the solid line in Figure 1 is obtained incorporating the inelastic-tunneling correction as in equation (19) at $A = 1.59 \times 10^{-9}$ s$^{-1}$ K$^{-1}$, while the remaining parameters $\varepsilon_B$, $\hbar\omega$ and $\eta$ are the same as the ones of the dashed-line fit therein. Figure 2 shows the extended thermally-activated range of the temperature dependencies, experimental & theoretical. The two fits of Figure 1, by dashed and solid lines, respectively, are very close to each other in the thermally-activated range. A nice concord is thus manifested between experiment and theory. Our Arrhenius branch is to be compared with earlier data in Figure 4.7 of Wert & Thomson's monograph [1].

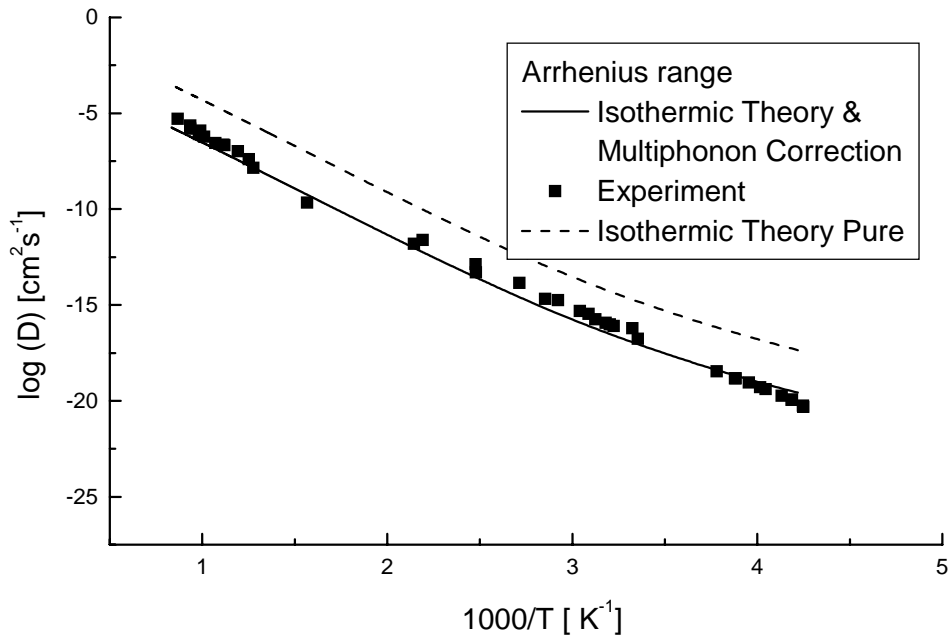

Figure 2. Temperature dependence of the carbon diffusion coefficient within the Arrhenius range from Reference [2÷4] experimental points by symbols. Our fits are: (i) isothermal reaction-rate fit (dashed) and (ii) fit combining basic isothermal calculations with one-phonon corrections (solid), as in Fig.1. The numerical values of $D$ are in cm$^2$s$^{-1}$. Compare with earlier diffusion data from Wert & Thomson's monograph [1]. $E_b = 1.1$ eV, $\hbar\omega = 0.077$ eV, $\eta = 0.025$.

### 3.3. Analysis

The obtained fitting and derivative parameters are listed next in Table I. One is the reorganization energy $\varepsilon_R = \frac{1}{2} K(\Delta Q_0)^2 \sim 4.626$ eV where $\Delta Q_0 = 2Q_0$ is the interwell separation. Next we calculate the stiffness $K = M\omega^2 = 17$ eV/Å$^2$ for $M = 12$ a.m.u. and $\hbar\omega = 0.077$ eV. Inserting we get $\Delta Q_0 \sim 0.738$ Å. (We note that $\Delta Q$ relating to configurational space, it may not be directly convertible to the jump distance d.) Another one is the electron-mode coupling strength G obtainable from the coupling energy $\varepsilon_{CE} = G^2/2K \sim 1.157$ eV: we get $G = \sqrt{(2E_{CE}K)} \sim 6.272$ eV/Å. From $\eta = V_{12}/2\varepsilon_{CE} \sim 0.025$ we now estimate $V_{12} \sim 0.058$ eV. Using the estimated parameters and setting E = 0, we show in Figure 3 the potential energy profile for carbon diffusion in α-iron, as obtained from equation (3).

Table I

Fitting and derivative parameters for the adiabatic potential energy surface controlling carbon diffusion in α-iron

| Fitting parameters | | | | | Derivative parameters | | | |
|---|---|---|---|---|---|---|---|---|
| $\varepsilon_B$ | $\hbar\omega$ | $\eta$ | $K$ | $G$ | $\varepsilon_{CE}$ | $\varepsilon_R$ | $\varepsilon_C$ | $\Delta Q$ |
| [eV] | [eV] | | [eV/Å$^2$] | [eV/Å] | [eV] | [eV] | [eV] | [Å] |
| 1.1 | 0.077 | 0.025 | 17 | 6.272 | 1.157 | 4.626 | 1.158 | 0.738 |
| $2V_{12}$ | $D(0)_{th}$† | $D(0)_{exp}$ | $d$ | $P = f(\zeta/6)d^2$ | $f$ | | $A$ | |
| [eV] | [cm$^2$/s] | [cm$^2$/s] | [Å] | [cm$^2$] | | | [s$^{-1}$K$^{-1}$] | |
| 0.116 | 3.7×10$^{-27}$ 1.2×10$^{-24}$ | 6.17×10$^{-25}$ | 0.74 | 6.50×10$^{-17}$ | 1.79 | | 1.59×10$^{-9}$ | |

†$D(0)_{thh} = P (\varepsilon_R / \hbar) \exp(-\varepsilon_R / \hbar\omega)$
$D(0)_{thv} = P (\Delta / k_B) A$

The obtained numerical values of K and G, essential as they are for choosing between various model predictions, seem at least reasonable in view of the high coupled vibrational frequency. But, model predictions for poorly conducting solids are less applicable to metals, due to screening by the electron gas. In spite of the uncertainties, there is no way of explaining the temperature dependence of the diffusion coefficient other than by means of the coupling to lattice vibrations. With specific differences between fairly and poorly conducting solids in mind, we also believe the present attempt may be found useful for dealing with impurity diffusion problems in metals.

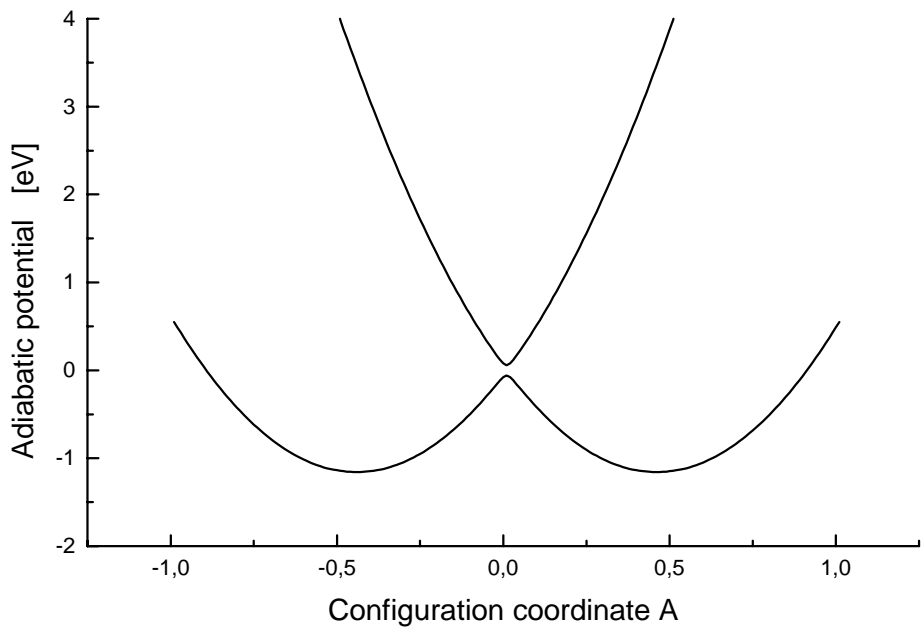

Figure 3. Calculated two-site adiabatic potential energy profile along the coupled-mode coordinate for carbon diffusion in α-iron using the parametric data in Fig.1 and Table 1.

We also compare the experimental scaling factor $P$ with the theoretical formula relating diffusion to rate as in equation (13) to get $f \sim 1$. At low temperature, therefore, the amount of carbon atoms making the efficient "forward jumps" in coherent migration is rather high.

## 4. Discussion

Undoubtedly, due to its practical significance, the system composed of carbon impurity atoms embedded in an iron frame has been one of the first to study scientifically in mankind's historic retrospective. It is amazing how long it has taken before quantal features of carbon impurity migration have been revealed even though somewhat accidentally.

In the foregoing work, we compared the temperature dependencies predicted for the bare rates, elastic and inelastic, to the experimental diffusion data on carbon in α–iron. From this comparison we obtained fitting values for the rate parameters, such as the migration hindering barrier and the coupled vibrational frequency. Both fitting values are in concert with the experimental estimates which manifests the applicability of the idealized bare rate to describing the carbon diffusion data. The correspondence between bare and observed rates suggests that dressing effects due to thermal bath coupling in the carbon & iron system may be of the text-book type. Indeed, the retardation due to thermal bath dressing comes from mass and coupled frequency renormalization. Among other things renormalization results in changes (reduction) of the effective nearest-neighbor jump distance (else the interwell separation) and of the effective tunneling frequency of carbon diffusion. As regards retardation caused by conduction electrons, we find no trace of its fingerprint in the experimental temperature dependence of the apparent rate within the tunneling range. We also find no evidence for any carbon-carbon interaction effects in diffusion at least at the low carbon densities involved (4% presently) which is not unexpected. In summary, we conclude that small-polaron theories seem to be an useful tool for dealing with the migration of carbon interstitials in α–iron.

The bare-rate theory used presently to tackle the diffusion problem is one that accounts for elastic-tunneling transitions mainly, though small corrections were made to include inelastic tunneling as well. Elastic tunneling leads to a coherent migration in which the number of phonons is conserved. The alternative is the theory accounting for inelastic-tunneling multiphonon transitions in which migration is accompanied by the absorption and emission of phonons. Accordingly, inelastic tunneling leads to a diffusion coefficient composed of coherent and incoherent components [7]. In the multiphonon model, the coherent tunneling is predicted dominating at the lowest temperatures though giving way to the incoherent process as the phonon exchange becomes important. For this reason our approach conserving the number of phonons should be comparable to the multiphonon theory at the lowest temperatures which makes desirable a parallel analysis of the multiphonon fundamentals. As a matter of fact, the multiphonon theory predicts that one-phonon processes with rate proportional to $T$ will follow suit as the temperature is raised. At the same time, the horizontal-tunneling likely prediction is for a constant rate until thermal transitions to the higher lying vibronic level become sizeable. The situation is envisaged by the experimental data in Figure 1 where the small increment of the measured diffusion coefficient data over the zero-point data below LNT is attributed to the one-phonon processes. A matter for further improvements of the accuracy of the isothermal rate approach is working out a better adiabatic potential, such as the trigonometric potential, which are now in progress.

Experimental evidence comes to show that the migrating carbon in α-iron occupies interstitial sites or pores. There are two types of an interstice in α-iron: less voluminous (00″) octahedral sites and more spacious (00Ľ) tetrahedral sites. It is a textbook assertion that carbon prefers the former sites despite their associated smaller volume. [14] On incorporating the impurity in an octahedral interstice, the local symmetry reduces from cubic to tetragonal. The controversy seems related to the lattice distortions produced by the incorporated atom: if in a tetra-pore C will have to displace all four nearby Fe atoms so as to make room for itself, while if in an octa-pore the induced deformation will spread to only two nearest neighbors.

One way or the other, the horizontal tunneling analysis is a powerful tool which provides, among other things, an estimate for the separation between neighboring sites in diffusion, and indeed, given that real- space and configurational- space coordinates run along similar straight lines, the site separation amounts to $\Delta Q_0$. From our estimates based on the experimental data, we get $\Delta Q_0 \sim \sqrt{2}/2 = 0.71$ Å which implies that the impurity jumps along the line segments connecting nearest-neighbor [½00] octahedral sites whose nn separation along the interconnecting segment is $(\sqrt{2}/2)$ a where a is the cubic lattice parameter. With a = 2.68 Å we see that the intersite separation would match our estimate for $\Delta Q_0$, provided both sites, initial and final, are displaced some 25-30 percent towards the tetrahedral positions along the vertical axis to make way for the impurity jump. In this sense our fitting results suggest that even though chiefly incorporated in octahedral sites carbon may prefer tetrahedral pores for migration across the α-iron lattice.

There apparently is more than one channel for carbon diffusion in iron, for example, o-o, o-t-o, etc. If more than one channel is operative, the present theory considers an effective rate process along a single effective configurational (mode) coordinate coupled to an effective vibrational frequency. Now, the actual configurational situation can be reconstructed with the help of additional considerations. In the present case the good agreement of the fitting vibrational frequency to the experimental frequency tells that our effective configurational coordinate is close to the actual one.

Another problem mentioned above is the likely symmetry of the coupled vibrational mode. We assume that it is the vibrations of the small carbon atom which drive it jumping from site to site. We obtain a fitting evidence confirming the identity of our frequency parameter (77 meV) with the carbon frequency obtained from inelastic neutron scattering (76 meV).[15] The estimate of *K* made above also assigns the whole vibrational feature to the carbon impurity. This implies a complete isotope effect $v \propto M^{-1/2}$ for the diffusing entity. Alternative symmetric models are also conceivable.